# A Survey of Link Recommendation for Social Networks: Methods, Theoretical Foundations, and Future Research Directions

Zhepeng Li,
Department of Operations and Information Systems
York University
zli@ schulich.yorku.ca

Xiao Fang*
Department of Management Information Systems
University of Delaware
xfang@udel.edu

Olivia Sheng
Department of Operations and Information Systems
University of Utah
Olivia.sheng@eccles.utah.edu

*Send comments to xfang@udel.edu

# A Survey of Link Recommendation for Social Networks: Methods, Theoretical Foundations, and Future Research Directions

**ABSTRACT**

Link recommendation has attracted significant attentions from both industry practitioners and academic researchers. In industry, link recommendation has become a standard and most important feature in online social networks, prominent examples of which include "People You May Know" on LinkedIn and "You May Know" on Google+. In academia, link recommendation has been and remains a highly active research area. This paper surveys state-of-the-art link recommendation methods, which can be broadly categorized into learning-based methods and proximity-based methods. We further identify social and economic theories, such as social interaction theory, that underlie these methods and explain from a theoretical perspective why a link recommendation method works. Finally, we propose to extend link recommendation research in several directions that include utility-based link recommendation, diversity of link recommendation, link recommendation from incomplete data, and experimental study of link recommendation.

**1. Introduction**

A social network can be represented as $G = <V, E>$, where $V$ denotes the set of social entities in the network and $E$ represents the set of existing links each of which connects a pair of social entities. For example, a social network on Facebook consists of a set of users (i.e., social entities) connected by pairwise friendship links among them. Let $\bar{E}$ be the set of potential links that have not been established in a social network; i.e., $\bar{E} = V \times V - E$. We use $F(\bar{e})$ to denote the value of a potential link $\bar{e} \in \bar{E}$. The link recommendation problem is to estimate the value of each potential link, rank potential links in decreasing order of value, and recommend top-ranked potential links.

Link recommendation has attracted significant attentions from both industry practitioners and academic researchers over the years. In industry, link recommendation has become a standard and most important feature in online social networks since its early success at LinkedIn (Davenport and Patil 2012).

Prominent examples of link recommendation include "People You May Know" on Facebook and LinkedIn as well as "You May Know" on Google+. In academia, link recommendation has been and remains a highly active research area. Given the tremendous academic and practical interests in link recommendation, there is a need for a review of the state-of-the-art knowledge in this area as well as the identification of significant and interesting questions for future research. This survey aims to address this need. Our survey differs from prior surveys of link prediction for social networks (Hasan and Zaki 2011, Lü and Zhou 2011) and contributes to the literature in the following ways:

(i) While prior surveys focus primarily on either learning-based link recommendation methods (Hasan and Zaki 2011) or proximity-based link recommendation methods (Lü and Zhou 2011), this survey reviews representative methods in both categories and is more comprehensive.

(ii) Prior surveys have not examined social and economic theories underlying link recommendation methods. Our survey identifies these theories (e.g., social interaction theory) and explains from a theoretical perspective why a link recommendation method works.

(iii) Our survey suggests a unique set of research directions worthy of future exploration.

In general, existing link recommendation methods operationalize $F(\bar{e})$ as the likelihood that potential link $\bar{e}$ will be established in the future and recommend the links that are most likely to be established (Kashima and Abe 2006, Hasan et al. 2006, Liben-Nowell and Kleinberg 2007, Yin et al. 2010, Backstrom and Leskovec 2011, Gong et al. 2012, Xia et al. 2012). Therefore, the core of these methods is the prediction of linkage likelihood. According to different prediction approaches used, existing link recommendation methods can be broadly categorized into learning-based methods and proximity-based methods, which we review in Sections 2 and 3 respectively. We summarize link recommendation methods in Section 4. We then identify social and economic theories underlying link recommendation methods in Section 5 and suggest important future research directions in Section 6. Section 7 concludes the paper.

## 2. Learning-based Methods

Learning-based methods learn a model from training data constructed from observed link establishments and use the learned model to predict the linkage likelihood for each potential link. A model can be learned using classification approaches, probabilistic models, or relational learning approaches. Learning-based methods can thus be categorized as classification-based methods, probabilistic model-based methods, and relational learning-based methods.

### 2.1 Classification-based Methods

Given a social network, we can construct training data from observed link establishments in the network. In general, each record of the training data has the format $< f_1, f_2, ..., f_m, l >$, where $f_1, f_2, ..., f_m$ represent features that affect link establishments in the network and $l$ is the class label. The class label $l$ is 1 for an existing link and it is 0 for a potential link. Commonly used features include topological features that are derived from the structure of a social network and nodal features that are computed from intrinsic characteristics of individual social entities. Topological features include neighbor-based features and path-based features. Neighbor-based features characterize neighborhoods of social entities, e.g., the number of common neighbors between social entities (Hasan et al. 2006, Lichtenwalter et al. 2010, Benchettara et al. 2010), while path-based features describe paths connecting social entities in a social network, e.g., the shortest distance between social entities (Hasan et al. 2006, Lichtenwalter et al. 2010, O'Madadhain et al. 2005, Benchettara et al. 2010, Wang et al. 2007). Nodal features can be computed from intrinsic characteristics of social entities, including demographic characteristics (Zheleva et al. 2008), geographic characteristics (O'Madadhain et al. 2005, Scellato et al. 2011, Wang et al. 2011), and semantic characteristics (Hasan et al. 2006, Wang et al. 2007).

Typical classification approaches can be applied to the constructed training data to predict the linkage likelihood of a potential link. O'Madadhain et al. (2005) employ logistic regression to predict the likelihood of interactions between social entities using data regarding CiteSeer articles, AT&T telephone calls, and Enron emails. Wang et al. (2007) combine a local Markov random field model and logistic regression to predict the likelihood of co-authorship using DBLP and PubMed article datasets. Benchettara et al. (2010) also predict the likelihood of co-authorship but employ a decision tree classifier enhanced with Adaboost for making the prediction. Gong et al. (2012) predict new and missing links in Google+ using support vector machine (SVM). Comparing popular classification approaches for predicting linkage likelihood, including Naïve Bayes, decision tree, SVM, and $k$-nearest neighbor, Hasan et al. (2006) find that SVM seems to be the most effective among these approaches and that SVM with a radial basis function (RBF) kernel outperforms SVM with a linear kernel.

The performance of classification-based methods can be enhanced in a number of ways. One way to enhance the effectiveness and efficiency of these methods is through judicious selection of relevant features. In this vein, Xu and Rockmore (2012) and Bliss et al. (2014) design feature selection frameworks for ranking and weighting features. In addition, Lichtenwalter et al. (2010) suggest that the performance of classification-based methods can be improved by carefully sampling training data. According to Lichtenwalter et al. (2010), to predict the likelihood of linkage between social entities that are $n$ hops away, it is more effective and efficient to use a training sample of entities that are $n$ hops apart than to use the entire training data. Yet another way to enhance performance is to enrich a social network with additional information that is useful for linkage likelihood prediction. A social network can be enriched by incorporating additional relationships among its social entities. For example, Zheleva et al. (2008) enrich a friendship social network of pets with their family affiliations. A social network can also be enriched by adding additional nodes. For instance, Gong et al. (2012) propose a social-attribute network whose nodes consist of both social entities and attributes of these entities. Finally, traditional ways of improving the performance of classification approaches are beneficial too. For example, ensemble methods, such as AdaBoost and Bagging, have been employed to improve the performance of linkage likelihood prediction (Benchettara et al. 2010, Hasan et al. 2006, Lichtenwalter et al. 2010), and Doppa et al. (2010) develop a cost-sensitive method to address the class imbalance issue in linkage likelihood prediction.

## 2.2 Probabilistic Model-based Methods

In general, probabilistic model-based methods (Kashima and Abe 2006, Clauset et al. 2008, Lu et al. 2010, Backstrom and Leskovec 2011, Hopcroft et al. 2011, Yang et al. 2011, Kim and Leskovec 2011, Barbieri et al. 2014, Dong et al. 2015, Song et al. 2015) predict the linkage likelihood $L_{ij}$ between social entities $v_i$ and $v_j$ as

$$L_{ij} = g(\boldsymbol{f}_{ij}, \boldsymbol{\theta}^*), \tag{1}$$

where $g(\cdot)$ is a prediction function, $\boldsymbol{f}_{ij}$ is a vector of features gathered from social entities $v_i$ and $v_j$, and $\boldsymbol{\theta}^*$ is a vector of parameters that can be learned from observed link establishments. Given training data $D$ constructed from observed link establishments, $\boldsymbol{\theta}^*$ is estimated as the parameter vector that best explains $D$ – that is,

$$\boldsymbol{\theta}^* = \underset{\boldsymbol{\theta}}{\mathrm{argmax}}\, Obj(D|M, \boldsymbol{\theta}), \tag{2}$$

where the objective function $Obj(D|M, \boldsymbol{\theta})$ defines how well observed link establishments can be explained by a probabilistic model $M$ with a parameter vector $\boldsymbol{\theta}$. We next discuss representative probabilistic model-based methods.

The method proposed by Kashima and Abe (2006) is based on the intuition that a social entity's linkage decision is influenced by his or her neighbors in a social network. Accordingly, they predict the linkage likelihood $\phi_{ij}^{t+1}$ between social entities $v_i$ and $v_j$ at time $t+1$ as

$$\phi_{ij}^{t+1} = \frac{1}{|V|-1}\left(\sum_{k \neq i,j} \theta_{kj}\phi_{ki}^t + \theta_{ki}\phi_{kj}^t\right) + \left(1 - \frac{1}{|V|-1}\sum_{k \neq i,j}\theta_{kj} + \theta_{ki}\right)\phi_{ij}^t, \tag{3}$$

where $V$ is the set of social entities in a social network, entity $k$ is a neighbor of entity $i$ or $j$, parameter vector $\boldsymbol{\theta} = <\theta_{xy}|\forall x, y \in V>$. The stationary likelihood of linkage between a pair of social entities is obtained by iteratively updating their linkage likelihood according to Equation (3) until convergence. The objective function $Obj(\cdot)$ is defined as the total stationary linkage likelihood summed over all existing links, and $\boldsymbol{\theta}^*$ is then estimated as the parameter vector that maximizes the objective function.

Guimerà and Marta (2009) propose a stochastic block model in which social entities are partitioned into groups. Given any two groups η and η′, there is a connection between them if there exists a link between entities $v_a \in η$ and $v_b \in η′$. Let $P_{ηη′}$ be the probability of a connection existing between groups η and η′. Let $A$ denote the adjacency matrix of a social network, where $A_{ab} = 1$ if there is a link between entities $v_a$ and $v_b$ in the network and $A_{ab} = 0$ otherwise. Given a partition model $M$, the likelihood of $A$ is given by

$$L(A|M) = \prod_{η \leq η′} P_{ηη′}^{l_{ηη′}} (1 - P_{ηη′})^{r_{ηη′} - l_{ηη′}}, \qquad (4)$$

where $l_{ηη′}$ is the number of existing connections between groups η and η′ and $r_{ηη′}$ is the number of possible connections between groups η and η′. Using Equation (4) as the objective function, we obtain the maximum likelihood estimation of $P_{ηη′}$ as

$$P_{ηη′}^* = \frac{l_{ηη′}}{r_{ηη′}}. \qquad (5)$$

The linkage likelihood $L_{ij}$ between social entities $v_i$ and $v_j$ is then given by

$$L_{ij} = \int_{\mathbf{M}} P(A_{ij} = 1|M) L(M|A) dM, \qquad (6)$$

where **M** is the space of all possible partition models. By applying Bayes' Theorem, we can rewrite Equation (6) as

$$L_{ij} = \frac{\int_{\mathbf{M}} P(A_{ij} = 1|M) L(A|M) P(M) dM}{\int_{\mathbf{M}} L(A|M) P(M) dM}, \qquad (7)$$

where $P(M)$ is the prior probability of partition model $M$ and $L(A|M)$ can be obtained using Equation (4) with $P_{ηη′}^*$. In practice, it is not possible to enumerate all possible partition models and the Metropolis algorithm is applied to sample partition models (Metropolis et al. 1953). Clauset et al. (2008) propose a similar method, in which social entities are grouped into a hierarchical structure.

Huang (2010) argues that the linkage likelihood between social entities $v_i$ and $v_j$ depends on the number of existing paths connecting them. Let $c_z$ be the probability of linking $v_i$ and $v_j$ given that there exists a length-$z$ path connecting them, $z \geq 2$. According to Huang (2010), the linkage likelihood $L_{ij}$ between social entities $v_i$ and $v_j$ can be predicted as

$$L_{ij} = 1 - (1-c_2)^{m_2}(1-c_3)^{m_3} \cdots (1-c_k)^{m_k}, \tag{8}$$

where $m_z$ denotes the number of existing length-$z$ paths connecting $v_i$ and $v_j$, $z = 2,3,\ldots,k$. To compute $L_{ij}$, we need to estimate parameters in $\boldsymbol{\theta} = <c_2, c_3, \ldots, c_k>$. Huang (2010) proposes to estimate $\boldsymbol{\theta}$ based on a generalized variant of the clustering coefficient, which indicates the tendency of a path to form a cycle (Watts and Strogatz 1998, Newman et al. 2001). Considering a social network as a graph, Huang (2010) computes its clustering coefficient $CC_k$ as

$$CC_k = \frac{|cycle^{<k>}|}{|path^{<k>}|}, \tag{9}$$

where $path^{<k>}$ denotes the set of length-$k$ paths and $cycle^{<k>}$ represents the set of cycles each of which is formed by adding a link to a length-$k$ path in $path^{<k>}$. By adding a link connecting $v_i$ and $v_j$ with probability $L_{ij}$, the expected clustering coefficient of the social network is a function of $\boldsymbol{\theta}$, i.e., $f(c_2, c_3, \ldots, c_k)$. According to Huang (2010), we have $f(c_2, c_3, \ldots, c_k) \approx CC_k$. Therefore, each parameter in $\boldsymbol{\theta}$ can be estimated as

$$c_z^* = \underset{c_z}{\mathrm{argmin}}(|f(c_2, c_3, \ldots, c_i, \ldots, c_k) - CC_k|), \tag{10}$$

where $z = 2,3,\ldots,k$.

Hopcroft et al. (2011) develop a method based on social network theories such as homophily theory. According to Hopcroft et al. (2011), the linkage likelihood $L_{ij}$ between social entities $v_i$ and $v_j$ is predicted as

$$L_{ij} = P(y_{ij} = 1 | \boldsymbol{f}_{ij}, \boldsymbol{\theta}^*), \tag{11}$$

where $y_{ij} = 1$ indicates the linkage between social entities $v_i$ and $v_j$, feature vector $\boldsymbol{f}_{ij}$ is constructed based on social network theories, and $\boldsymbol{\theta}^*$ is the parameter vector to be learned. Specifically, $\boldsymbol{\theta}^*$ is estimated as the parameter vector that best explains observed link establishments using a gradient descent method (Hopcroft et al. 2011).

Backstrom and Leskovec (2011) propose a supervised random walk method to predict linkage likelihood. They define a transition matrix with the social entities in a social network as indices. Each element of the transition matrix is a transition probability from one entity to another and it is defined as an exponential function

$$a_{ij} = \exp(\boldsymbol{f}_{ij} \cdot \boldsymbol{\theta}^*), \tag{12}$$

or a logistic function

$$a_{ij} = \frac{1}{1 + \exp(-\boldsymbol{f}_{ij} \cdot \boldsymbol{\theta}^*)}, \tag{13}$$

where $\boldsymbol{f}_{ij}$ is a feature vector. To compute the linkage likelihood $L_{ij}$ between entities $v_i$ and $v_j$, a random walk with restart is started from $v_i$ according to the transition matrix and $L_{ij}$ is the stationary probability of reaching $v_j$ from $v_i$. Backstrom and Leskovec (2011) estimate $\boldsymbol{\theta}^*$ as the parameter vector that minimizes a loss function defined based on the difference between the stationary probability from $v_i$ to one of its currently linked social entities and the stationary probability from $v_i$ to one of the social entities that it is currently not linked to. Yang et al. (2011) develop a similar linkage prediction method that combines a random walk model with collaborative filtering techniques.

**2.3 Relational Learning-based Methods**

Relational learning represents data using a relational model; it then learns and infers based on the model (Friedman et al. 1999, Getoor et al. 2001, Taskar et al. 2003, Heckerman et al. 2004). Specifically, a relational model consists of two components: relational data model ($M_D$) and dependency structure model ($M_S$). A relational data model consists of a set of classes $\boldsymbol{X} = \{X^c\}$, each of which is described by a set of descriptive attributes $\boldsymbol{A}(X^c)$ (Getoor and Taskar 2007). For example, as shown in Figure 1, PAPER class $X^{paper}$ has descriptive attributes such as topic area, keyword, and journal, i.e., $\boldsymbol{A}(X^{paper}) = \{Area, Keyword, Journal\}$. Each class $X^c$ is associated with a set of linkage relationships $\boldsymbol{R}(X^c)$. In Figure 1, a linkage relationship $x^{paper}.\rho \in R(X^{paper})$ of a paper $x^{paper} \in X^{paper}$ represents its citation of other papers.

PAPER Class

| PaperID | Area | Keyword | Journal |
|---|---|---|---|
| P1 | AI | Representation | J1 |
| P2 | Data Mining | Clustering | J2 |
| P3 | Data Mining | Association Rule | J1 |
| P4 | AI | Reasoning | J1 |

Citation

| Citing | Cited |
|---|---|
| P1 | P4 |
| P2 | P3 |
| P3 | P1 |

Figure 1: An Example of Relational Data Model ($M_D$)

A dependency structure model represents the dependencies among the descriptive attributes and linkage relationships of a class. Using the PAPER class in Figure 1 as an example, a dependency structure model shown in Figure 2 captures the following dependencies: (a) the area of a paper depends on the journal in which it is published, (b) the keyword of a paper depends on its area, and (c) linkage relationships (e.g., citations) among papers depends on their areas, keywords, and published journals. Dependencies can be learned from data or constructed using background knowledge (Getoor and Taskar 2007).

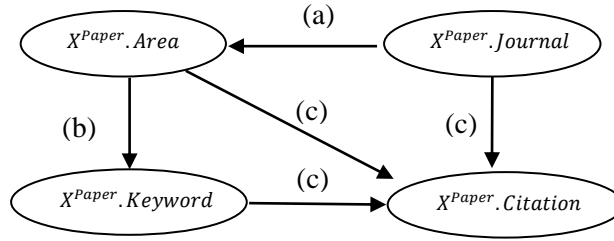

Figure 2: An Example of Dependency Structure Model ($M_S$)

Given a dependency structure model, the distribution of a descriptive attribute $x^c.A$ and the distribution of a linkage relationship $x^c.\rho$ can be derived as $P(x^c.A|Pa(x^c.A))$ and $P(x^c.\rho|pa(x^c.\rho)$ respectively, where $Pa(\cdot)$ denotes the parents of a descriptive attribute or a linkage relationship in a dependency structure model. The objective of relational learning is to obtain a vector of parameters $\boldsymbol{\theta}$ such that the following data likelihood function for an instance $\boldsymbol{I}$ of a relational model is maximized.

$$l(\boldsymbol{\theta}|M_D, M_S, \boldsymbol{I}) = \prod_{X^c \in \boldsymbol{X}} \prod_{x^c \in X^c} \prod_{A \in \boldsymbol{A}(X^c)} P(x^c.A|Pa(x^c.A)) \prod_{\rho \in \boldsymbol{R}(X^c)} P(x^c.\rho|Pa(x^c.\rho)). \qquad (14)$$

The obtained optimal parameters $\boldsymbol{\theta}^*$ can then be used to infer the probability of a linkage relationship (e.g., citation) and the distribution of a missing descriptive attribute (Getoor et al. 2003).

To apply relational learning to the link recommendation problem, we can treat linkage likelihood in link recommendation as the probability of a linkage relationship in relational learning (Getoor et al. 2003). In this vein, Bilgic et al. (2007) show that inferring the distributions of missing descriptive attributes can help the estimation of linkage likelihood. In addition to Bayesian Network-based relational learning depicted in Equation (14), there are other relational learning methods that can be applied to link recommendation. For example, Markov Network-based relational learning has been used for link recommendation (Taskar et al. 2003, Mihalkova et al. 2011). Moreover, Heckerman et al. (2001) introduce a relational dependency network for relational learning, Popescul and Ungar (2003) develop a relational learning method based on the structural logistic regression model, Heckerman et al. (2004) design a relational learning model based on the Entity-Relationship diagram, and Yu et al. (2007) propose a non-parametric stochastic approach to relational learning.

## 3. Proximity-based Methods

Proximity-based methods surrogate the linkage likelihood between social entities using the proximity between them (Liben-Nowell and Kleinberg 2007, Chen et al. 2009, Crandall et al. 2010, Lü and Zhou 2011). According to homophily theory (McPherson et al. 2001), similar social entities are of high tendency to link to each other. In light of this theory, the greater the proximity between social entities the higher the linkage likelihood between them. Proximity-based methods can generally be grouped into nodal proximity-based methods and structural proximity-based methods.

### 3.1 Nodal Proximity-based Methods

Nodal proximity-based methods surrogate the linkage likelihood $L_{ij}$ between social entities $v_i$ and $v_j$ using their nodal proximity $S(\boldsymbol{Y}_i, \boldsymbol{Y}_j)$, where $\boldsymbol{Y}_i$ and $\boldsymbol{Y}_j$ denote the profile of social entities $v_i$ and $v_j$ respectively and $S(\cdot)$ is a similarity function. The profile of a social entity consists of its intrinsic characteristics, including demographic, geographic, and semantic characteristics. Demographic characteristics, such as age, education, and occupation, are commonly used (Zheleva et al. 2008, Xu and Rockmore 2012). Geographic characteristics, such as co-location and distance, have also been employed to capture the closeness between two social entities in a physical space (Quercia and Capra 2009, Crandall et al. 2010, Yin et al. 2010, Scellato et al. 2011, Wang et al. 2011). Semantic characteristics, such as keywords, annotation tags, and communication descriptions, are used to measure the similarity between social entities in terms of their semantic patterns (Shen et al. 2006, Chen et al. 2009, Schifanella et al. 2010, Yin et al. 2010, Makrehchi 2011, Adali et al. 2012, Kuo et al. 2013, Yuan et al. 2014).

A number of similarity functions have been applied to measure nodal proximity. For profiles with numerical characteristics, cosine similarity has been employed (Chen et al. 2009, Shen et al. 2006, Schifanella et al. 2010, Wang et al. 2011). According to Salton (1989), the cosine similarity between social entities $v_i$ and $v_j$ is given by

$$CS(\mathbf{Y}_i, \mathbf{Y}_j) = \frac{\sum_k Y_{ik} Y_{jk}}{\sqrt{\sum_k Y_{ik}^2 \sum_k Y_{jk}^2}}, \tag{15}$$

where $Y_{ik}$ is the $k^{th}$ characteristic in $\mathbf{Y}_i$. Another similarity function suitable for numerical characteristics is KL-divergence (Shen et al. 2006). Concretely, the KL-divergence between social entities $v_i$ and $v_j$ is computed as

$$KL(\mathbf{Y}_i, \mathbf{Y}_j) = \sum_k h_{ik} \log \frac{h_{ik}}{h_{jk}} + h_{jk} \log \frac{h_{jk}}{h_{ik}}, \tag{16}$$

where $h_{ik}$ is the probability of the $k^{th}$ characteristic in $\mathbf{Y}_i$. Manhattan distance has been used to gauge the *dissimilarity* between social entities (Wang et al. 2011, Adali et al. 2012). Specifically, the Manhattan distance between social entities $v_i$ and $v_j$ is defined as

$$MD(\mathbf{Y}_i, \mathbf{Y}_j) = \sum_k |Y_{ik} - Y_{jk}|. \tag{17}$$

For profiles with nominal characteristics, Jaccard coefficient is a suitable similarity function (Schifanella et al. 2010, Scellato et al. 2011, Wang et al. 2011, Xu and Rockmore 2012, Kuo et al. 2013). In particular, the Jaccard coefficient between social entities $v_i$ and $v_j$ is defined as

$$JC(Y_i, Y_j) = \frac{|Y_i \cap Y_j|}{|Y_i \cup Y_j|}. \tag{18}$$

In addition to the above-reviewed similarity functions, other similarity functions such as match count have also been used (Kahanda and Neville 2009, Xiang et al. 2010).

### 3.2 Structural Proximity-based Methods

Structural proximity measures the proximity between two social entities based on their structural features in a social network (Jeh and Widom 2002, Liben-Nowell and Kleinberg 2007, Lü et al. 2009, Lü and Zhou 2011, Liu and Lü 2010). Structural proximity-based methods surrogate the linkage likelihood $L_{ij}$ between social entities $v_i$ and $v_j$ with their structural proximity (Liben-Nowell and Kleinberg 2007), which can be classified into neighborhood-based structural proximity and path-based structural proximity.

### 3.2.1 Neighborhood-based Structural Proximity

Structural proximity between social entities can be measured based on their neighborhoods. Common neighbor is a widely used neighborhood-based structural proximity measure (Newman 2001, Liben-Nowell and Kleinberg 2007). The common neighbor $CN_{ij}$ between social entities $v_i$ and $v_j$ is computed as the number of their mutual neighbors, i.e.,

$$CN_{ij} = |\Gamma_i \cap \Gamma_j|, \tag{19}$$

where $\Gamma_i$ and $\Gamma_j$ denote the set of direct neighbors of entities $v_i$ and $v_j$ respectively and $|\cdot|$ is the cardinality of a set. Extended from the common neighbor measure, the Adamic/Adar measure assigns less weight to more connected common neighbors (Adamic and Adar 2003, Liben-Nowell and Kleinberg 2007). Specifically, the Adamic/Adar $AA_{ij}$ between social entities $v_i$ and $v_j$ is given by

$$AA_{ij} = \sum_{v_z \in \Gamma_i \cap \Gamma_j} \frac{1}{\log |\Gamma_z|}, \tag{20}$$

where $v_z$ is a common neighbor of $v_i$ and $v_j$ and $\Gamma_z$ denotes the set of direct neighbors of $v_z$. It has been shown theoretically and empirically that the linkage likelihood between social entities is highly correlated with their neighborhood sizes (Barabási and Albert 1999, Newman 2001, Barabási et al. 2002, Liben-Nowell and Kleinberg 2007). Motivated by these theoretical and empirical findings, the preferential attachment $PA_{ij}$ between social entities $v_i$ and $v_j$ is defined as

$$PA_{ij} = |\Gamma_i| \times |\Gamma_j|. \tag{21}$$

Observing that two social entities are similar if their neighbors are similar, Jeh and Widom (2002) propose the SimRank measure. The SimRank score $SR_{ij}$ between social entities $v_i$ and $v_j$ is defined as (Jeh and Widom 2002, Liben-Nowell and Kleinberg 2007, Liu and Lü 2010)

$$SR_{ij} = \frac{\gamma \cdot \sum_{v_z \in \Gamma_i} \sum_{v_{z'} \in \Gamma_j} SR_{zz'}}{|\Gamma_i| \cdot |\Gamma_j|}, \tag{22}$$

where $\gamma \in (0,1)$ is a decay factor. A variation of the SimRank measure is proposed by Leicht et al. (2006). In addition to the measures reviewed in this subsection, there other neighborhood-based measures, such as the Sørensen Index (Sørensen 1948), the Salton Measure (Salton and McGill 1986), and the Hub Promoted (HP)/Hub Depressed (HD) Index (Ravasz et al. 2002).

### 3.2.2 Path-based Structural Proximity

Going beyond neighborhoods, path-based structural proximity measures target paths connecting social entities. The Katz index measures the structural proximity between social entities using the number of paths connecting them, weighted by their lengths (Katz 1953). Originally developed for measuring the social status of a social entity, the Katz index has been shown to be effective in predicting linkage between social entities (Liben-Nowell and Kleinberg 2007). The Katz index $KZ_{ij}$ between social entities $v_i$ and $v_j$ is given by (Katz 1953)

$$KZ_{ij} = \sum_k \beta^k \left| path_{ij}^{<k>} \right|, \tag{23}$$

where $path_{ij}^{<k>}$ represents the set of length-$k$ paths connecting $v_i$ and $v_j$ and weight $\beta$ is between 0 and 1. According to Equation (23), the contribution of a path to the Katz index decreases as its length increases. The local path index (Zhou et al. 2009) is a localized version of the Katz index, where $k$ in Equation (23) is upper bounded.

Considering link establishment between social entities as a random walk from one to the other, the PageRank algorithm (Brin and Page 1998) can be adapted to compute structural proximity between them (Haveliwala 2002, Jeh and Widom 2003, Tong et al. 2006). Let $\boldsymbol{P}$ be a transition matrix with element $P_{xy}$ representing the transition probability from social entity $v_x$ to entity $v_y$, where $P_{xy} = \frac{1}{|\Gamma_x|}$ if $e_{xy} \in E$ and $P_{xy} = 0$ otherwise. Let $\boldsymbol{q}_i$ be a vector of probabilities, each element of which represents the probability from social entity $v_i$ to another entity in a social network. For example, element $q_i^j$ represents the probability from $v_i$ to $v_j$. According to Tong et al. (2006), we have

$$\boldsymbol{q}_i = \alpha \boldsymbol{P}^T \boldsymbol{q}_i + (1-\alpha)\boldsymbol{\epsilon}_i, \tag{24}$$

where the $i^{th}$ element in vector $\boldsymbol{\epsilon}_i$ is 1 and all other elements are 0. We can iteratively compute $\boldsymbol{q}_i$ according to Equation (24) until convergence. The PageRank-based structural proximity $PR_{ij}$ between $v_i$ and $v_j$ is calculated as

$$PR_{ij} = q_i^j + q_j^i. \tag{25}$$

Based on a similar idea of treating link establishment as a random walk, Fouss et al. (2007) define the hitting time $H_{ij}$ as the expected number of steps needed to reach social entity $v_j$ from entity $v_i$ for the first time. Let $b_{xy}$ denote the weight between entities $v_x$ and $v_y$, where $b_{xy} > 0$ if $v_x$ and $v_y$ are linked and $b_{xy} = 0$ otherwise. To reach $v_j$ from $v_i$, one needs to go to a neighbor $v_k$ of $v_i$ and then proceeds from $v_k$ to $v_j$. Accordingly, the hitting time $H_{ij}$ can be measured as

$$H_{ij} = 1 + \sum_k P_{ik} H_{kj}. \tag{26}$$

where $P_{ik}$ is the probability of moving from $v_i$ to its neighbor $v_k$ and $P_{ik} = b_{ik}/\sum_z b_{iz}$. The hitting time $H_{ij}$ can be computed recursively according to Equation (26) using a Markovian algorithm (Kemeny and Snell 1976) or can be obtained using the pseudoinverse of the Laplacian matrix of a social network (Fouss et al. 2007). The average commute time $ACT_{ij}$ is a symmetric variation of $H_{ij}$. Specifically, the average commute time $ACT_{ij}$ is defined as the expected number of steps needed to reach $v_j$ from $v_i$ for the first time and then to go back to $v_i$ from $v_j$ (Fouss et al. 2007). Thus, $ACT_{ij}$ is measured as (Fouss et al. 2007)

$$ACT_{ij} = H_{ij} + H_{ji}. \tag{27}$$

Liben-Nowell and Kleinberg (2007) define the normalized average commute time $ACT'_{ij}$ as

$$ACT'_{ij} = H_{ij} \cdot \pi_j + H_{ji} \cdot \pi_i, \tag{28}$$

where $\pi_j$ and $\pi_i$ represent the stationary probability of reaching $v_j$ and $v_i$ respectively. Since $H_{ij}$, $ACT_{ij}$, and $ACT'_{ij}$ are all distance-based measures, a higher value of these measures indicates less linkage likelihood (Liben-Nowell and Kleinberg 2007).

### 3.2.3 Performance Improvement

Several approaches have been proposed to improve the performance of structural proximity-based methods. A social network can be represented as an adjacency matrix, each element of which represents whether two social entities are linked or the strength of their linkage. Inspired by the success of matrix factorization techniques in information retrieval and recommender systems (Deerwester et al. 1990, Koren et al. 2009, Rennie and Srebro 2005), Liben-Nowell and Kleinberg (2007) apply singular value decomposition, a matrix factorization technique, to reduce the rank of an adjacency matrix that represents a social network. They show that structural proximities computed using the rank-reduced adjacency matrix can predict linkage more accurately than those calculated from the original adjacency matrix. Other approaches to improving the prediction accuracy of structural proximity-based methods include the aggregation approach (Liben-Nowell and Kleinberg 2007) and the kernel approach (Kunegis and Lommatzsch 2009).

Given the huge size of real world social networks, it is imperative to develop efficient approaches to computing structural proximities. One approach (Song et al. 2009, Shin et al. 2012) is developed based on approximation techniques. It trades accuracy for efficiency and produces approximate structural proximities. Another approach calculates structural proximities from a local social network rather than a complete social network and thus improves the efficiency of computing structural proximities (Zhou et al. 2009, Lü et al. 2009, Lichtenwalter et al. 2010, Liu and Lü 2010, Fire et al. 2011). Yet another improvement adds a time dimension to a social network and computes structural proximities from an evolving social network. For example, Huang and Lin (2009) combine time series models and structural proximity measures to predict linkage likelihood at a particular time. Acar et al. (2009) and Dunlavy et al. (2011) create a tensor from a series of adjacency matrices, each of which represents a social network at a particular time; they then compute the Katz index from the tensor.

**4. Summary of Link Recommendation Methods**

In summary, link recommendation methods predict the linkage likelihood for each potential link and recommend potential links with the highest linkage likelihoods. In particular, these methods predict linkage likelihood using machine learning approaches, including classification, probabilistic models, and relational learning, or they surrogate linkage likelihood with proximity measures, such as nodal and structural proximity measures. We therefore categorize link recommendation methods accordingly and list representative works in each category in Table 1.

|  | Method | Representative Work |
|---|---|---|
| Learning-based Method | Classification-based Method | - O'Madadhain et al. 2005<br>- Hasan et al. 2006<br>- Wang et al. 2007<br>- Lichtenwalter et al. 2010<br>- Scellato et al. 2011<br>- Gong et al. 2012<br>- Bliss et al. 2014<br>- Zhang et al. 2014 |
| | Probabilistic Model-based Method | - Kashima and Abe 2006<br>- Guimerà and Marta 2009<br>- Huang 2010<br>- Backstrom and Leskovec 2011<br>- Hopcroft et al. 2011<br>- Yang et al. 2011<br>- Barbieri et al. 2014<br>- Dong et al. 2015 |
| | Relational Learning-based Method | - Getoor et al. 2003<br>- Popescul and Ungar 2003<br>- Taskar et al. 2003<br>- Yu et al. 2006<br>- Mihalkova et al. 2011 |
| Proximity-based Method | Nodal Proximity-based Method | - Chen et al. 2009<br>- Crandall et al. 2010<br>- Schifanella et al. 2010 |
| | Structural Proximity-based Method (Neighborhood) | - Newman 2001<br>- Barabási et al. 2002<br>- Jeh and Widom 2002<br>- Adamic and Adar 2003<br>- Liben-Nowell and Kleinberg 2007 |
| | Structural Proximity-based Method (Path) | - Tong et al. 2006<br>- Liben-Nowell and Kleinberg 2007 |

**Table 1:** Categorization of Link Recommendation Methods

In comparison to proximity-based methods, learning-based methods have the following advantages. First, the output of a learning-based method is (or can be easily transformed to) the probability that a potential link will be established in the future. For example, a Naïve Bayes-based method outputs the probability of linkage for each potential link. Most proximity-based methods, on the other hand, do not produce probabilities of linkage. Taking nodal proximity-based methods as an example, these methods yield similarity scores between social entities rather than probabilities of linkage. While the outputs of proximity-based methods are adequate for ranking potential links, they are insufficient for more advanced applications that require probabilities of linkage. For example, in many applications, the cost of recommending a wrong potential link that will not be established is different from the cost of missing a true potential link that will be established. In these applications, knowing probabilities of linkage is essential to cost-sensitive link recommendation decisions. Second, well-designed learning-based methods can predict future linkage more accurately than proximity-based methods (Lichtenwalter et al. 2010, Backstrom and Leskovec 2011). This is partly due to the fact that learning-based methods learn to predict future linkage from the ground truth of prior link establishments (Lichtenwalter et al. 2010). Another factor contributes to the advanced prediction performance of learning-based methods is that their prediction performance is stable across different social networks, while the prediction performance of proximity-based methods varies from one network to another (Lichtenwalter et al. 2010). Proximity-based methods, however, have their own merits. First, proximity-based methods do not need a training phase, which is required for learning-based methods. As a consequence, compared to learning-based methods, proximity-based methods save the cost of constructing training data and the cost of learning from training data. Second, many proximity-based methods are easy to implement and widely applied in practice. For example, common neighbor is popularly used by major online social networks for link recommendation. It is called "mutual friend" in Facebook and "shared connection" in LinkedIn.

## 5. Theoretical Foundations for Link Recommendation Methods

Our survey of link recommendation methods suggests that many of these methods are developed based on observations of social phenomena but overlook social and economic theories underlying these phenomena. As we illustrate in Figure 3, the link between link recommendation methods and their foundational theories is largely missing in existing link recommendation studies. In general, a theory explains why a mechanism works and under what circumstances it works (Parsons 1938, Bacharach 1989). In particular, identifying social and economic theories underlying link recommendation methods has the following benefits. First, it helps us understand why and under what situations a link recommendation method works. With this understanding, we can effectively select appropriate methods for link recommendation tasks at hand. Second, it helps us identify limitations of existing methods and design more advanced methods. Third, theories inform us about generic factors that actually affect linkage decisions, which we can use to design novel and effective link recommendation methods. In this section, we identify social and economic theories underlying link recommendation methods and uncover the missing link between them.

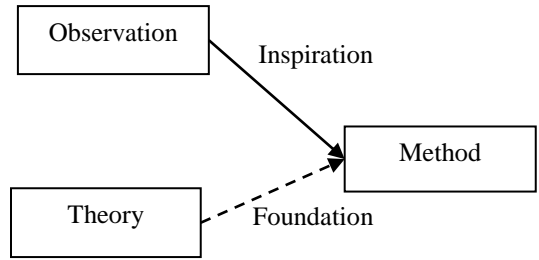

**Figure 3:** The Missing Link between Social/Economic Theories and Link Recommendation Methods

Observations that have inspired existing link recommendation methods can be broadly grouped into three categories. Specifically, it is observed that the decision of linkage between two social entities is affected by (1) the degree of similarity between them (O'Madadhain et al. 2005, Hasan et al. 2006, Wang et al. 2007, Chen et al. 2009, Crandall et al. 2010, Schifanella et al. 2010, Mihalkova et al. 2011, Hopcroft et al. 2011, Scellato et al. 2011, Yang et al. 2011, Gong et al. 2012), (2) linkage decisions of their social neighbors (Newman 2001, Barabási et al. 2002, Jeh and Widom 2002, Adamic and Adar 2003, O'Madadhain et al. 2005, Kashima and Abe 2006, Lichtenwalter et al. 2010, Scellato et al. 2011, Gong et al. 2012), and (3) paths connecting them in a social network (Katz 1953, Hasan et al. 2006, Tong et al. 2006, Wang et al. 2007, Huang 2010, Lichtenwalter et al. 2010, Hopcroft et al. 2011, Gong et al. 2012). These observations can be explained by homophily theory (McPherson et al. 2001), social interaction theory (Becker 1974), and cognitive balance theory (Heider 1958) respectively. We next review these theories and explain why a link recommendation method works based on these theories.

Homophily theory states that "a contact between similar people occurs at a higher rate than among dissimilar people" (McPherson et al. 2001). After all, people like to hang out with others who are like them; i.e., birds of a feather flock together. Theoretically, the flow of information from one person to another is a decreasing function of the distance between them in the Blau space, which is defined by socio-demographic dimensions (McPherson 1983). As a consequence, people separated by greater distances in the Blau space (i.e., less similar in terms of socio-demographics) are less likely to be linked (McPherson 1983). Furthermore, people self-select their "social world" by choosing with whom to interact. Similarity localizes communications in a social network and can lead to distinct social niches (McPherson and Ranger-Moore 1991). These self-selection effects and social niches imply that similarity contributes to the establishment of linkage in social networks (McPherson and Ranger-Moore 1991).

Homophily theory is the theoretical foundation for a number of link recommendation methods. According to this theory, a person tends to link to another person similar to him or her and the linkage between them further strengthens their similarity (McPherson and Ranger-Moore 1991). This mutual reinforcement between similarity and linkage is the foundation for the link recommendation methods developed by Yang et al. (2011) and Gong (2012). Moreover, homophily theory predicts that "similarity breeds connection" (McPherson et al. 2001), which explains the effectiveness of nodal proximity-based methods (Chen et al. 2009, Crandall et al. 2010, Schifanella et al. 2010). Furthermore, using similarities between user profiles, one can build a Markov network for relational learning-based link recommendation (Mihalkova et al. 2011). In addition, homophily theory also justifies the effectiveness of similarity-based input features for link recommendation methods, including semantic similarities (O'Madadhain et al. 2005, Hasan et al. 2006, Wang et al. 2007), geographic proximities (O'Madadhain et al. 2005, Scellato et al. 2011), similarities on personal interests (Yang et al. 2011), social status similarities (Hopcroft et al. 2011), and demographic similarities (Gong et al. 2012).

According to social interaction theory (Becker 1974), when making a social decision, a social entity's utility depends on the decisions of his or her social neighbors. Specifically, a social decision is a decision that has impacts on the decision maker's social neighbors (Akerlof 1997). For link recommendation, the decision of linkage is a social decision. Similar to social interaction theory, social decision theory states that the welfare of a decision maker is affected by the outcomes of his or her social neighbors (Akerlof 1997, Maccheroni et al. 2012). Both social interaction theory and social decision theory can be reasoned from the perspective of social information processing (Salancik and Pfeffer 1978). That is, a social entity relies on social information, i.e., information about what others think, to make a social decision (Kelley 1971, Berger and Calabrese 1975, Pfeffer et al. 1976, Salancik and Pfeffer 1978). In particular, a social entity's decision is affected by the decisions of his or her social neighbors in two ways (Salancik and Pfeffer 1978). First, the entity tends to fit into a social environment by agreeing with his or her social neighbors. Second, the entity's social neighbors can pass selective information to the entity, thereby affecting the entity's decision.

In short, social interaction theory, social decision theory, and social information processing theory all suggest that a social entity's decision depends on the decisions of the entity's social neighbors, which is the theoretical foundation for a group of link recommendation methods. For example, Kashima and Abe (2006) develop a link recommendation method based on the idea that a user's linkage decision in a social network depends on the linkage decisions of his or her direct friends in the network. Moreover, the influence of a social entity's neighborhood on the entity's decision, as suggested by these theories, is also encoded in the development of neighborhood-based structural proximities, which function as independent predictors for link recommendation (Newman 2001, Barabási et al. 2002, Jeh and Widom 2002, Adamic and Adar 2003) or as input features for a link recommendation method (O'Madadhain et al. 2005, Scellato et al. 2011, Gong et al. 2012).

Network transitivity distinguishes social networks from other types of networks (Newman and Park 2003). In general, network transitivity refers to the social phenomenon that indirectly associated social entities tend to tie to each other directly (Davis and Leinhardt 1971, Huang 2010). This phenomenon can be explained by cognitive balance theory from a psychological perspective (Heider 1958) or by social focus theory from a sociological standpoint (Homans 1950). According to cognitive balance theory, sentiments (or attitudes) of indirectly associated social entities could become consistent gradually, which in turn could drive them to link to each other (Heider 1958). Social focus theory suggests that people connect with those who share their foci (Homans 1950). Here, a focus is defined as "a social, psychological, legal, or physical entity around which joint activities are organized" (Feld 1981). For example, foci can be workplaces or hangouts. For two indirectly associated social entities, intermediate entities connecting them provide or create foci for them to meet and interact; as a consequence, these indirectly associated social entities are likely to be linked to each other (Feld 1981). Cognitive balance theory and social focus theory are theoretical foundations for link recommendation methods using path-based structural proximities (Katz 1953, Tong et al. 2006) and the probabilistic model-based method proposed by Huang (2010). All of these methods are developed based on network transitivity suggested by these theories. In addition, features constructed based on network transitivity have been used as inputs for several classification-based link recommendation methods (Hasan et al. 2006, Wang et al. 2007, Lichtenwalter et al. 2010, Gong et al. 2012) and the probabilistic model-based link recommendation method proposed by Hopcroft et al. (2011).

In summary, we identify major social and economic theories underlying link recommendation methods and explain working mechanisms of link recommendation methods using these theories, thereby uncovering the link between link recommendation methods and their theoretical foundations. Table 2 summarizes these theories and representative link recommendation methods that are grounded in the theories.

| Theory | Representative Work Grounded in Theory |
|---|---|
| Homophily Theory (McPherson et al. 2001) | Nodal Proximity-based Method<br>    - Chen et al. 2009<br>    - Crandall et al. 2010<br>    - Schifanella et al. 2010<br>Classification-based Method<br>    - O'Madadhain et al. 2005<br>    - Hasan et al. 2006<br>    - Wang et al. 2007<br>    - Scellato et al. 2011<br>    - Gong et al. 2012<br>Probabilistic Model-based Method<br>    - Hopcroft et al. 2011<br>    - Yang et al. 2011<br>Relational Learning-based Method<br>    - Mihalkova et al. 2011 |
| Social Interaction Theory (Becker 1974)<br><br>Social Information Processing Theory (Salancik and Pfeffer 1978)<br><br>Social Decision Theory (Akerlof 1997) | Probabilistic Model-based Method<br>    - Kashima and Abe 2006<br>Structural Proximity-based Method (Neighborhood)<br>    - Newman 2001<br>    - Barabási et al. 2002<br>    - Jeh and Widom 2002<br>    - Adamic and Adar 2003<br>Classification-based Method<br>    - O'Madadhain et al. 2005<br>    - Lichtenwalter et al. 2010<br>    - Scellato et al. 2011<br>    - Gong et al. 2012 |
| Social Focus Theory (Homans 1950)<br><br>Cognitive Balance Theory (Heider 1958) | Structural Proximity-based Method (Path)<br>    - Tong et al. 2006<br>Probabilistic Model-based Method<br>    - Huang 2010<br>    - Hopcroft et al. 2011<br>Classification-based Method<br>    - Hasan et al. 2006<br>    - Wang et al. 2007<br>    - Lichtenwalter et al. 2010<br>    - Gong et al. 2012 |

**Table 2:** Theoretical Foundations for Link Recommendation Methods

## 6. Future Research Directions

We propose to extend current link recommendation research in several directions, including utility-based link recommendation, diversity of link recommendation, link recommendation from incomplete data, and experimental study of link recommendation. In the following subsections, we identify promising research questions for each of these directions.

**6.1 Utility-based Link Recommendation**

Existing link recommendation methods predict the linkage likelihood for each potential link and recommend potential links with the highest linkage likelihoods. Thus, existing methods focus on the accuracy of link recommendation. In addition to accuracy, there are other factors that need to be considered when recommending links. One is the value of link recommendation; i.e., the value of a recommended link if it is established. We illustrate the value of link recommendation using an example of Facebook, whose operator (i.e., Facebook Inc.) harvests the majority of its $7.9 billion revenue from advertisements on the network (Facebook 10-K 2013). Facebook allows an advertisement being placed on the Facebook page of selected users. A Facebook user could interact with the advertisement through actions such as click, comment, like, and share. Such interactions propagate the advertisement to the user's friends, who could also interact with the advertisement and further propagate it to their friends. As this propagation process continues, the advertisement could reach a much larger number of users than those that were initially selected. Facebook Inc. reaps revenue each time the advertisement reaches a Facebook user. Understandably, a recommended link, if established, leads to a more connected network among Facebook users, which in turn, drives advertisements to reach more users and bring more advertisement revenue to Facebook Inc. In this example, the value of link recommendation is the advertisement revenue brought in by a recommended and established link. Additionally, link recommendation is not costless. For example, in the context of online social network, ineffective link recommendations suggest new friends that are mostly irrelevant for the users. Consequently, new friendships recommended may not be established, and users may feel disappointed and eventually leave the network. Due to the failure of establishing new friendships and the loss of users, an online social network becomes smaller and less connected, resulting in less advertisement revenue being produced. Hence, another critical factor for link recommendation is the cost of link recommendation; i.e., the cost incurred if a recommended link is not established.

We propose the utility of link recommendation that integrates the accuracy, value, and cost of link recommendation. Specifically, the utility of a recommended link depends on its linkage likelihood, the value brought in by the link if it is established, and the cost incurred if it is not established. An interesting research question is therefore how to recommend potential links with the highest utilities rather than the highest linkage likelihoods. This research question could be framed as a classification problem with the linkage likelihood, value, and cost of a potential link as predictors (Li et al. 2015). Alternatively, this question could be treated as a cost-sensitive learning problem (Fang 2013), in which, the linkage likelihood of a potential link is predicted using an existing link recommendation method and a cost matrix is constructed based on the value and cost of a potential link.

**6.2 Diversity of Link Recommendation**

Besides accuracy and utility, another important objective of link recommendation is diversity, e.g., recommending friends with diverse backgrounds to a user. Diverse link recommendation benefits both individual users and a social network as a whole. A user who receives friend recommendations with diverse backgrounds could gain accesses to different social communities in a social network (Brandão et al. 2013), thereby obtaining significant social benefits such as a variety of information sources and competitive advantages (Burt 1992). A social network with high structural diversity among its users effectively facilitates the diffusion of information over the network (Ugander et al. 2012). Moreover, such a network is robust in terms of information diffusion efficiency, even after the removal of its well-connected users (Albert et al. 2000, Tanizawa et al. 2005). In addition, according to Eagle et al. (2010), high structural diversity in a social network has been shown to be positively correlated with a high level of socioeconomic well-being of its users (e.g., high income).

Despite these benefits, link recommendation diversity, however, has not received much research attention. In this subsection, we propose several research questions in this fruitful area. First, it is necessary to design metrics to gauge the diversity of link recommendation. While a number of measures has been proposed to evaluate the recommendation diversity of recommender systems (e.g., Zhang and Hurley 2008, Vargas and Castells 2011), many of these measures are not readily applicable to measuring the diversity of link recommendation because a recommender system is a user-item bipartite graph whereas link recommendation involves a user-user graph. We therefore suggest some ideas for designing metrics for link recommendation diversity. One possible metric would be the number of connected components among friends recommended to a user. Understandably, a larger number of connected components indicates that recommended friends are distributed in more social groups and hence greater link recommendation diversity. We could also evaluate link recommendation diversity by measuring the diversity of a social network after adding recommended links with classical network metrics such as clustering coefficient (Watts and Strogatz 1998), network diameter (Wasserman and Faust 1994), and number of structural holes (Burt 1992). Second, it would be interesting to empirically evaluate the recommendation diversity of popular link recommendation methods using the designed metrics. Third, methods that balance multiple objectives of link recommendation, such as diversity, utility, and accuracy, are needed. In this vein, we could formulate a problem of maximizing the diversity of link recommendation while maintaining the accuracy (or utility) of link recommendation at a certain level or a problem of maximizing the accuracy (or utility) of link recommendation while maintaining the diversity of link recommendation at a certain level, and develop methods to solve these problems.

**6.3 Link Recommendation from Incomplete Data**

Current link recommendation methods employ observed features of a social network to predict linkage likelihood. Some of these target nodal features (e.g., Backstrom and Leskovec 2011), while others use structural features (e.g., Liben-Nowell and Kleinberg 2007). However, unobserved factors for which we do not have data could also drive link establishments in a social network, in addition to observed features. Consider the following examples:

- Parents become friends on Facebook because their children attend the same kindergarten. These parents have quite different profiles and would never become friends if their children were not attending the same kindergarten.

- A group of researchers who have different backgrounds and are located in geographically separated areas are temporarily assigned to work together on an interdisciplinary project. Some of them then link to each other on LinkedIn.

The factors driving the link establishments in these examples are children's attendance at the same kindergarten and temporary project assignments. We do not have data about these factors (i.e., they are unobserved), partly for the following reasons. First, users may not disclose sensitive information such as which kindergarten their children attend due to privacy concerns. Second, users live in a social environment that is much broader than a social network. As a result, many factors extrinsic to a social network, such as temporary project assignments, could drive link establishments in the network. The sheer number of extrinsic factors that might affect link establishments in a social network make it extremely difficult to collect data about all of them. Existing methods that rely on similarities between observed user profiles fail to predict link establishments in these examples because observed user profiles in these examples are quite dissimilar. Hence, unobserved factors are important for link recommendation. However, we do not have data about unobserved factors, so our data for predicting linkage likelihood are incomplete. Therefore, an interesting research question is how to integrate both observed features and unobserved factors for the purpose of link recommendation. Toward this end, methods built on the expectation-maximization framework (Dempster et al. 1977, Fang et al. 2013a), a classical framework for learning from incomplete data, are needed for link recommendation from incomplete data.

**6.4 Experimental Study of Link Recommendation**

Most link recommendation methods are evaluated using archival data. However, simply using archival data, we cannot differentiate between link establishments due to organic growth and link establishments arising from link recommendation. Furthermore, with archival data alone, it is difficult to assess users' behavioral and emotional reactions to recommended links. Therefore, it is worthwhile to conduct laboratory or field experiments to evaluate link recommendation methods. In an experiment, we can observe in real time how users react to recommended links, both behaviorally and emotionally, which recommended links they actually establish, and the value of the established links along with the costs of links that are not established. Moreover, we could use laboratory or field experiments to study the delivery mechanisms for link recommendation. Interesting questions in this direction include: How should the user interface that presents recommended links be designed? What are the best timing and frequency for recommending links that will maximize the chance of link establishment? How many links should be recommended to a user? And what incentives can be used to facilitate the establishment of recommended links? In addition, we could design experiments to study the impact of link recommendation on social networks. One interesting question in this direction is to study how link recommendation affects the diffusion of information, opinion, or advertisements in a social network. In short, experimental study provides a means to address interesting link recommendation questions, many of which cannot be answered using archival data alone. By combining laboratory or field experiments and evaluations with archival data, we could produce more comprehensive and convincing evaluations of link recommendation methods.

**6.5 Other Future Research Questions**

There are other interesting future research questions, which we discuss briefly for length consideration. The effectiveness of a link recommendation method decreases over time, because the method was developed based on previous user linkage behaviors and does not capture new user linkage behaviors. Thus, an interesting question is how to maintain the currency of a link recommendation method over time. Prior studies on data stream mining (Aggarwal 2007) and knowledge refreshing (Fang et al. 2013b) provide methodological and theoretical foundations for answering this question. In addition, current link recommendation methods rank potential links by their linkage likelihoods and recommend top-ranked links. Thus, existing methods essentially solve a ranking problem. However, many real world link recommendation situations may be modeled as combinatorial rather than ranking problems. For example, how to recommend a set of $K$ potential links that will collectively bring in the largest advertisement revenue, among all possible combinations of $K$ potential links, is a combinatorial problem. Thus, it is interesting to study how to model and solve combinatorial link recommendation problems. There are also other challenges for link recommendation, such as cold start (Leroy et al. 2010), class imbalance (Rattigan and Jensen 2005, Lichtenwalter et al. 2010), link recommendation in signed social networks (Song et al. 2015, Tang et al. 2015), and link recommendation across multiple (heterogeneous) social networks (Zhang et al. 2014, Dong et al. 2015), which are well documented in the literature and hence are not discussed here.

## 7. Conclusions

We review state-of-the-art link recommendation methods in this survey. We also identify social and economic theories underlying link recommendation methods and uncover the missing link between them. We further suggest several extensions to existing link recommendation studies, including utility-based link recommendation, diversity of link recommendation, link recommendation from incomplete data, and experimental study of link recommendation. We hope this survey serves as a useful summary of what has been done so far as well as a stimulus to the advancement of link recommendation research.